\documentclass[usenatbib,twocolumn]{mn2e}
\usepackage{times}
\usepackage{graphicx}
\date{\today}
\def\mnras{MNRAS}
\def\apj{ApJ}
\def\apjs{ApJ Supplement Series}
\def\apjl{ApJ Letters}
\def\physrep{Phys Reps}
\def\aj{A J}
\def\aap{A\&A}
\def\nat{Nature}
\def\prd{Phys. Rev. D}
\author[Yadav, Bagla and Khandai]{Jaswant K. Yadav$^{1,2}$, J. S. Bagla$^2$ and Nishikanta  Khandai$^3$\\
$^1$Korea Institute for Advanced Study, Hoegiro 87, Dongdaemun-gu, Seoul 130722,
South Korea\\
$^2$Harish-Chandra Research Institute,  Chhatnag Road, Jhusi,
Allahabad 211019, INDIA\\
$^3$McWilliams Center for Cosmology, Carnegie Mellon University, Pittsburgh, PA 15213, U.S.A. \\
{E-Mail: jaswant@kias.re.kr, jasjeet@hri.res.in, nkhandai@andrew.cmu.edu}}

\title[The scale of homogeneity]{Fractal Dimension as a measure of the scale of 
  Homogeneity}

\begin{document}

\maketitle 
\begin{abstract}
In the multi-fractal analysis of large scale matter distribution, the scale 
of transition to homogeneity is defined as the scale above which the fractal 
dimension ($D_q$) of underlying point distribution is {\it equal to} the 
ambient dimension (D) of the space in which points are distributed. 
With finite sized weakly clustered distribution of tracers obtained from
galaxy redshift surveys it is difficult to achieve this equality.  
Recently \citet{2008MNRAS.390..829B} have defined the scale of homogeneity to
be the scale above which the deviation ($\Delta D_q$) of fractal dimension
from the ambient dimension becomes smaller than the statistical dispersion of
$\Delta D_q$, i.e., $\sigma_{\Delta D_q}$. 
In this paper we use the relation between the fractal dimensions and the
correlation function to compute $\sigma_{\Delta D_q}$ for any given model in
the limit of weak clustering amplitude.   
We compare $\Delta D_q$ and $\sigma_{\Delta D_q}$ for the $\Lambda$CDM  model
and discuss the implication of this comparison for the expected scale of
homogeneity in the concordant model of cosmology. 
We estimate the upper limit to the scale of homogeneity to be close to
$260$~h$^{-1}$Mpc for the $\Lambda$CDM model. 
Actual estimates of the scale of homogeneity should be smaller than this as we
have considered only statistical contribution to $\sigma_{\Delta D_q}$ and we
have ignored cosmic variance and contributions due to survey geometry and the
selection function. 
Errors arising due to these factors enhance $\sigma_{\Delta D_q}$ and as
$\Delta D_q$ decreases with increasing scale, we expect to {\it measure} a
smaller scale of homogeneity.  
We find that as long as non linear correction to the computation of $\Delta
D_q$  are insignificant, scale of homogeneity does not change with epoch.   
The scale of homogeneity depends very weakly on the choice of tracer of the
density field.
Thus the suggested definition of the scale of homogeneity is fairly robust.
\end{abstract}


\section{Introduction}

One of the primary aims of galaxy redshift surveys is to determine the
distribution of luminous matter in the Universe
\citep { 1983ApJS...52...89H, 
  1995ApJ...455...50L, 1996ApJ...470..172S, 1999ApJS..121..287H, 
  2000AJ....120.1579Y,  2001MNRAS.328.1039C, 2004ApJ...600L..93G, 
  2005A&A...439..877L, 2007ApJS..172....1S}.
These surveys have revealed a large variety of structures starting
from groups and clusters of galaxies, extending to super-clusters and an
interconnected network of filaments which appears to extend across very large
scales \citep{1996Natur.380..603B, 2005Natur.435..629S, 2008Sci...319...52F}. 
We expect the galaxy distribution to be homogeneous on large scales. 
In fact, the assumption of large scale homogeneity and isotropy of the
universe is the basis of most cosmological models \citep{1917SPAW.......142E}.
In addition to determining the large scale structures, the redshift surveys of
galaxies can also be used to verify whether the galaxy distribution does
indeed become homogeneous at some scale\citep{1993ApJ...407..443E,
  1998MNRAS.298.1212M, 2005MNRAS.364..601Y, 2009EL.....8529002S,
  2009EL....8649001S}.    
Fractal dimensions of the galaxy matter distribution can be used to test the
conjecture of homogeneity. 
One advantage of using fractal dimensions over other analyses is that in
the former we don't require the assumption of an average density in the point
set \citep{2004RvMP...76.1211J}. 

When doing our analysis we often work with volume limited sub-samples
extracted from the the full magnitude samples of the galaxies.  
This is done in order to avoid an explicit use of the selection function.  
The volume limited sub-samples constructed in this manner from a flux limited
parent sample naturally have a much smaller number of galaxies.
This was found to be too restrictive for the earliest surveys and corrections
for varying selection function were used explicitly in order to determine the
scale of homogeneity \citep{1999A&A...351..405B, 1999ApJ...514L...1A}. 
But with modern galaxy redshift surveys, this limitation is less severe. 

Making a volume limited sub-sample requires assumption of a cosmological model
and this may be thought of as an undesirable feature of data analysis. 
However, if we directly use raw data and do not account for a redshift
dependent selection function in any way, it is obvious that the selection
function will dominate in any large scale description of the distribution of
galaxies.  
This is only to be expected as we see only brighter galaxies at larger
distances in a flux or magnitude limited sample. 

For a given survey one computes the fractal dimension of the point
set under consideration and the scale beyond which the fractal dimension is
equal to the ambient dimension of the space in which the particles are
distributed is referred to as the scale of homogeneity of that distribution. 
Mathematically the fractal dimension is defined for an infinite set of
points. 
Given that the observational samples are finite there is a need to understand
the deviations in the fractal dimensions arising due to the finite number of
points.  
In practice we should identify a scale to be the scale of homogeneity where
the {\it rms} error in the fractal dimension is comparable to or greater than
the deviation of the fractal dimension from the ambient dimension
\citep{2008MNRAS.390..829B}. 
It is not possible to distinguish between a given point set and a homogeneous
point set with the same number of points in the same volume beyond the scale
of homogeneity . 

In all practical cases of interest the fractal dimension is never equal to the
ambient integer dimension of the space.  
\citet{2008MNRAS.390..829B} have shown that these deviations in fractal
dimension occur mainly due to weak clustering present in the galaxy
distribution, with a smaller contribution arising due to the finite number of
galaxies in the distribution.   
This work assumed the standard cosmological model in order to derive a
relation between the fractal dimension on one hand, and, a combination 
of number density of tracers of the density field and clustering on the 
other. 
It was assumed that the clustering is weak at scales of interest.  
In this paper, we revisit the expected deviation of fractal dimension for
a finite distribution of weakly clustered points and verify the relations
derived in \citet{2008MNRAS.390..829B}. 
For this purpose we have used a particle distribution from a large volume
$N-$Body simulations.  
Further, we generalise the relation between clustering and fractal dimension
to compute the {\it rms} error in determination of fractal dimensions and
clustering from data. 
We then discuss the implications of this for the expected scale of homogeneity
in the standard cosmological model.  
We also comment on some recent determinations of the scale of homogeneity from
observations of galaxy distribution in redshift surveys.  

The plan of the paper is as follows. In section \S{2} we present a brief
introduction to fractals and fractal dimensions. Subsection \S{2.1} describes 
the fractal dimension for a weakly clustered distribution of finite number of
points. Section \S{3} discusses the calculation of variance in $\xi$ and hence
the variance in offset to fractal dimension.  We present the results in section
\S{4}, and conclude with a summary of the paper in \S{5}.

\section{Fractals and Fractal Dimensions}

The name {\it fractal} was introduced by Benoit B. Mandelbrot 
\citep{ 1982fgn..book.....M} to characterise geometrical figures which may not 
be smooth or regular. One of the definitions of a fractal is that it is a shape
made of parts similar to the whole in some way. It is useful to regard a 
fractal as a set of points that has properties such as those listed below, 
rather than to look for a more precise definition which will almost certainly 
exclude some interesting cases. A set $F$ is a fractal if it satisfies most of 
the following \citep{2003fgmfa.book.....P}:
\begin{enumerate}
\item 
$F$ has a fine structure, i.e., detail on arbitrarily small scales.
\item 
$F$ is too irregular to be described in traditional geometrical language, both
locally and globally. 
\item 
$F$ has some form of self-similarity, perhaps approximate or statistical. 
\item 
In most cases of interest $F$ is defined in a very simple way, perhaps
recursively. 
\end{enumerate}
Fractals are characterised using the so called fractal dimensions.  
These can be defined in different ways, which do not necessarily 
coincide with one another. 
Therefore, an important aspect of studying a fractal structure is the choice
of a definition for fractal dimension that best applies to the case in study. 

Fractal dimension provides a description of how much space a point 
set fills. 
It is also a measure of the prominence of the irregularities of a point set
when viewed at a given scale. 
We may regard the dimension as an index of complexity. 
We can, in principle, use the concept of fractal dimensions to characterise
any point set. 

The simplest definition of the fractal dimension is the so called {\it Box
  counting dimension}. 
Here we place a number of mutually exclusive boxes that cover the region in
space containing the point set and count the number of boxes that contain some
of the points of the fractal.   
The fraction of non-empty boxes clearly depends on the size of boxes.
Box counting dimension of a fractal distribution is defined in terms of non
empty boxes $N(r)$ of radius $r$ required to cover the distribution. 
If
\begin{equation}
 N(r) \propto r^{D_b}
\label{eq:4}
\end{equation}
we define $D_b$ to be the box counting dimension \citep{1995fcsg.book.....B}.
In general $D_b$ is a function of scale.  
One of the difficulties with such a definition is that it does not depend on
the number of particles inside the boxes and rather depends only on the number
of boxes. 
It provides very limited information about the degree of clumpiness of the
distribution and is more of a  geometrical measure. 
To get more detailed information on clustering of the distribution we use the
concept of correlation dimension. 
We have chosen the correlation dimension, among various other definitions
\citep[see e.g.][]{1995PhR...251....1B, 2002sgd..book.....M}, due to its
mathematical simplicity and the ease with which it can be adapted to
calculations for a given point set.  
The formal definition of correlation dimension demands that the number of
particles in the distribution should be infinite (e.g. $N\to\infty$ in
equation \ref{eq:5}). 
We use a working definition for a finite point set. 
Calculation of the correlation dimension requires the introduction of
correlation integral given by 
\begin{equation}
C_2(r)=\frac{1}{NM}\sum_{i=1}^{M}n_i(r)
\label{eq:5}
\end{equation}
Here we assume that we have $N$ points in the distribution. 
$M$ is the number of centres on which spheres of radius $r$ have been placed
with the requirement that the entire sphere lies inside the distribution of
points.  
Spheres are centred on points within the point set and it is clear that  $M <
N$.   
Here $n_i(r)$ denotes the number of particles within a distance $r$ from
$i^{th}$ point: 
\begin{equation}
n_i(r)=\sum_{j=1}^{N}\Theta(r-\mid x_i-x_j \mid)
\label{eq:6}
\end{equation}
where  $\Theta(x)$ is the  Heaviside function. 
If we consider the distribution function for the number of points inside such
spherical cells, we can rewrite $C_2$ in terms of the probability of finding
$n$ particles in a sphere of radius $r$. 
\begin{equation}
C_2(r)=\frac{1}{N}\sum\limits_{n=0}^{N}n P(n;r,N)
\label{eq:7}
\end{equation}
where $P(n;r,N)$ is the normalised probability of getting $n$ out of $N$
points as neighbours inside a radius $r$ of any of the points.  
The correlation dimension $D_2$ of the distribution of points can be defined
via the power law scaling of correlation integral, i.e., $C_2(r) \propto
r^{D_2}$. 
\begin{equation}
D_2(r)=\frac{\partial \log C_2(r)}{\partial \log r}
\label{eq:8}
\end{equation}
Since the scaling behaviour of $C_2$ can be different at different scales,
we expect the correlation dimension to be a function of scale.
For the special case of a homogeneous distribution we see that the correlation 
dimension equals the ambient dimension for an infinite set of points.

$C_2(r)$ defined in the manner given by equation \ref{eq:7} provides the
average of the distribution function of $P(n;r,N)$. 
In order to characterise the distribution of points, we need information about
all the moments of the distribution function. 
This leads us to the generalised dimension $D_q$, also known as
Minkowski-Bouligand dimension.  
This is a generalisation of the correlation dimension $D_2$.
The correlation integral can be generalised to define $C_q(r)$ as
\begin{equation}
C_q(r) = \frac{1}{N}\sum\limits_{n=0}^{N}n^{q-1} P(n) \equiv
\frac{1}{N}\langle {\mathcal{N}}^{q-1}\rangle_p 
\label{eq:9}
\end{equation}
which is used to define the Minkowski-Bouligand dimension
\begin{equation}
D_q(r)=\frac{1}{q-1}\frac{d\log{C_q(r)}}{d\log{r}}
\label{eq:10}
\end{equation}
We expect the generalised dimension to be scale dependent in general. 
If the value of $D_q(r)$ is independent of both $q$ as well as $r$ then the
point distribution is called a mono-fractal. 

If we are dealing with a finite number of points in a finite volume then we
can always define an average density. 
This allows us to relate the generalised correlation integral with correlation
functions.  
For $q > 1$, the generalised correlation integral and the Minkowski-Bouligand
dimension can be related to a combination of $n$-point correlation functions
with the highest $n$ equal to $q$ and the smallest $n$ equal to $2$.
The contribution to $C_q$ is dominated by regions of higher number density of
points for $q \gg 1$ whereas for $q\ll 0$ the contribution is dominated by
regions of very low number density. 
This clearly implies that the full spectrum of generalised dimension gives us
information about the entire distribution: regions containing clusters of points
as well as voids that have few points. 

This allows us to connect the concept of fractal dimensions with the
statistical measures used to quantify the distribution of matter at large
scales in the universe.  
In the situation where the galaxy distribution is homogeneous and isotropic on
large scales, we intuitively expect $D_q \simeq D = 3$ independent of the value
of $q$. 
Whereas at smaller scales we expect to see a spectrum of values for the
fractal dimension, all different from $3$. 
It is of considerable interest to find out the scale where we can consider the
universe to be homogeneous.  
We address this issue using theoretical models in this paper.

In an earlier paper we have derived a leading order expression for generalised
correlation integral for homogeneous as well as weakly clustered
distributions of points \citep{2008MNRAS.390..829B}. 
We found that even for a homogeneous distribution of finite number of points
the Minkowski-Bouligand dimension, $D_q(r)$, is not exactly equal to the
ambient dimension, $D$. 
The difference in these two quantities arises due to discreteness effects. 
Hence for a finite sample of points, the correct benchmark is not $D$ but the
estimated value of $D_q$ for a homogeneous distribution of same number of
points in the same volume.  
An interesting aside is that the correction due to a finite size sample always
leads to a smaller value for $D_q(r)$ than $D$. 
As expected, this correction is small if the average number of points in
spheres is large, i.e.,  $\bar{N} \gg 1$. 

\citet{2008MNRAS.390..829B} have demonstrated that the general expression 
for $m^{th}$ order moment of a weakly clustered distribution of points is given 
by: 
\begin{eqnarray}
  \langle{\mathcal{N}}^m\rangle_p &=& {\bar{N}}^{m}\left[ 1 +
  \frac{\left(m\right)\left(m-1\right)}{2{\bar{N}}} +
  \frac{m(m+1)}{2}\bar{\xi} \right. \nonumber \\ 
&&\left.+ {\mathcal{O}}\left({\bar\xi}^2\right) +
  {\mathcal{O}}\left(\frac{\bar\xi}{\bar{N}}\right)
  + {\mathcal{O}}\left(\frac{1}{{\bar{N}}^2}\right)\right]
  \label{eq34}
\end{eqnarray} 
where 
\begin{equation}
\bar N = nV \qquad \& \qquad \bar\xi(r) = \frac{3}{r^3} \int\limits_0^r
\xi(x)  x^2 dx  
\label{nxi}
\end{equation}  
is the average number of particles in a randomly placed sphere and the volume 
averaged two point correlation function respectively.

The generalised correlation integral (eq :\ref{eq:9}) for this distribution
can now be written as 
\begin{eqnarray}
  NC_q(r) &=&  \langle{\mathcal{N}}^{q-1}\rangle_p \nonumber \\
  & =& {\bar{N}}^{q-1}\left[ 1 +
  \frac{\left(q-1\right)\left(q-2\right)}{2{\bar{N}}} +
  \frac{q(q-1)}{2}\bar{\xi} \right. \nonumber \\ 
&&\left.+ {\mathcal{O}}\left({\bar\xi}^2\right) +
  {\mathcal{O}}\left(\frac{\bar\xi}{\bar{N}}\right)
  + {\mathcal{O}}\left(\frac{1}{{\bar{N}}^2}\right)\right] 
  \label{eqcqclus} 
\end{eqnarray} 
The third term on the right hand side in fact encapsulates the contribution of
clustering. 
This differs from the last term in the corresponding expression
for a homogeneous distribution as in that case the ``clustering'' is only due
to cells being centred at points whereas in this case the locations of every
pair of points has a weak correlation. 
It is worth noting that the highest order term of order
$\mathcal{O}\left({\bar\xi}^2\right)$ has a factor
$\mathcal{O}\left(q^3\right)$ and hence can become important for sufficiently
large $q$. 
This may be quantified by stating that $q\bar\xi \ll 1$ is the more relevant
small parameter for this expansion.
The Minkowski-Bouligand dimension for this system can be expressed as:
\begin{eqnarray}
D_q(r)&\simeq& D\left(1-\frac{(q-2)}{2\bar{N}}\right) +\frac{q}{2}\frac{\partial
   \bar \xi}{\partial \log r} \nonumber \\
&=& D\left(1-\frac{(q-2)}{2\bar{N}}\right) +\frac{qr}{2}\frac{\partial
   \bar \xi}{\partial r} \nonumber \\
&=& D\left(1-\frac{(q-2)}{2\bar{N}} - \frac {q}{2}\left(\bar\xi(r)
 -\xi(r) \right)\right) \nonumber \\
&=& D - \left(\Delta{D_q}\right)_{\bar{N}} -
  \left(\Delta{D_q}\right)_{clus} \nonumber \\
\Delta D_q(r) &=& - \left(\Delta{D_q}\right)_{\bar{N}} -
  \left(\Delta{D_q}\right)_{clus}  
\label{eqdqclus}
\end{eqnarray} 
For a weakly clustered distribution we note that
\begin{itemize}
\item 
For hierarchical clustering, both terms have the same sign and lead to a
smaller value for $D_q$ as compared to $D$. 
\item
Unless the correlation function has a feature at some scale, smaller
correlation corresponds to a smaller correction to the Minkowski-Bouligand
dimension.
\item
If the correlation function has a feature then it is possible to have a small
correction term $\left(\Delta{D_q}\right)_{clus}$ for a relatively large
$\xi$. 
In such a situation, the relation between $\xi$ and
$\left(\Delta{D_q}\right)_{clus}$ is no longer one to one. 
In such a case $\left(\Delta{D_q}\right)_{clus}$ does not vary monotonically
with scale.
\end{itemize}

The model described here has been validated with the multinomial-multi-fractal 
model \citep[see e.g.][]{2008MNRAS.390..829B}. 
In the following discussion, we validate our model with the help of a large
volume N-Body simulation in order to check it in the setting where we wish 
to use it. 

\begin{figure}
\begin{center}
\includegraphics[width=3.2in]{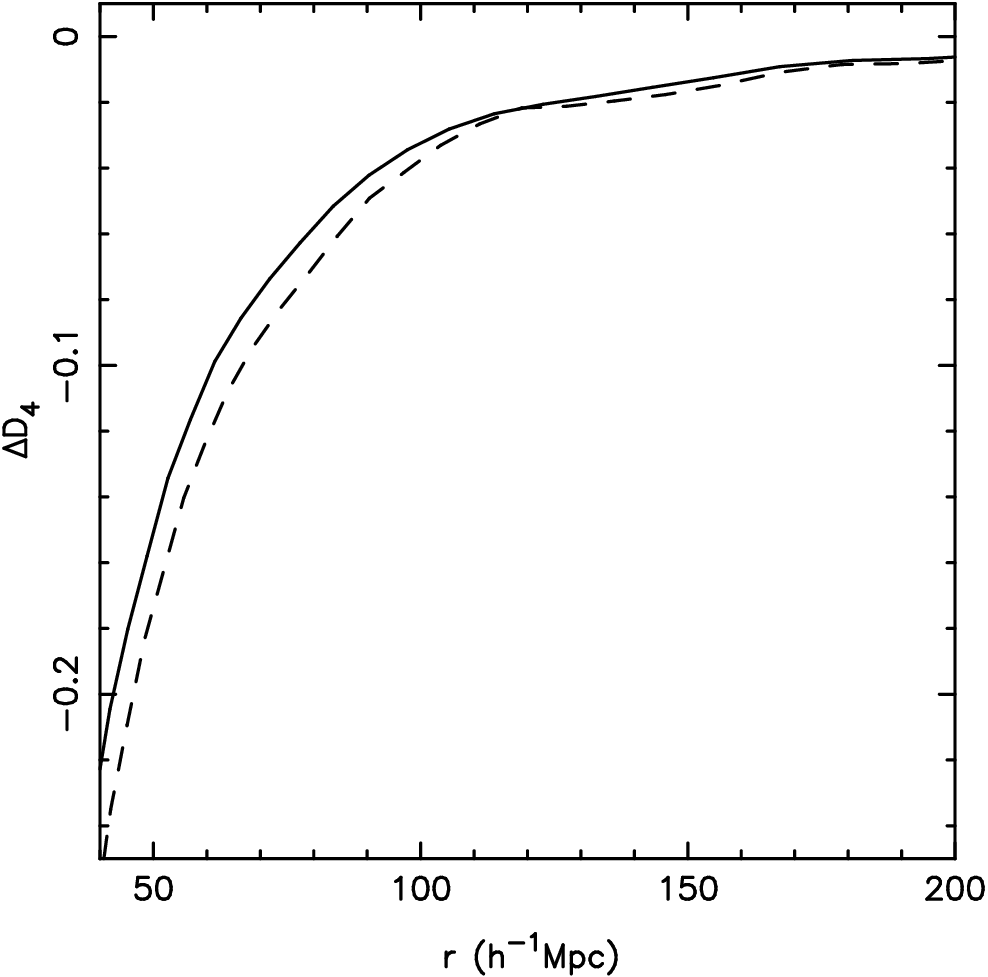}
\caption{Comparison of $\Delta D_4$ calculated using equation \ref{eq:10}
  (solid line), and estimated using equation \ref{eqdqclus}(dashed line) for
  the large volume $N-$Body simulation.} 
  \label{fig:dqvsr}
\end{center}
\end{figure}

\subsection{$N$-Body Simulations}

We use a the TreePM code for cosmological simulations
\citep{2002JApA...23..185B, 2003NewA....8..665B, 2009RAA.....9..861K} 
to simulate the distribution of particles. 
The simulations were run with the set of cosmological parameters favoured by
WMAP5 \citep{2009ApJS..180..330K} as the best fit for the $\Lambda$CDM class
of models: $\Omega_{nr} = 0.2565$, $\Omega_{\Lambda} =0.7435 $, $n_s = 0.963$,
$\sigma_8 = 0.796$, $h=0.719$, and, $\Omega_b h^2 = 0.02273$. 
The simulations were done with $512^3$ particles in a comoving cube of side
$1024 h^{-1} Mpc$.  

We computed the two point correlation function $\xi(r)$ directly from the 
$N-$Body simulation output by using a subsample of particles
\citep{1980lssu.book.....P, 1984ApJ...284L...9K}.  
The volume averaged correlation function $\bar{\xi}(r)$ follows from $\xi(r)$
using equation \ref{nxi}. 

Figure \ref{fig:dqvsr} presents the comparison between $\Delta D_q$ estimated 
directly using equation \ref{eq:10}, from the $N-$Body simulation, and, the
values computed using equation \ref{eqdqclus}.
For the last curve we use the values of $\xi(r)$ and $\bar\xi(r)$ computed
from the simulation.   
We find that the two curves track each other and the differences are less than
$10\%$ at all scales larger than $60$~h$^{-1}$Mpc. 
This is fairly impressive given that we have only taken the leading order
contributions into account.  
This validates the relation between the correlation functions and the fractal
dimensions in the limit of weak clustering. 
We find that as we go to larger scales, $\Delta D_q$ becomes smaller but does
not vanish. 
One may then ask, is there no scale where the universe becomes homogeneous? 
The answer to this question lies in a comparison of the offset 
$\Delta D_q$ with the dispersion expected due to statistical errors, we
discuss this in detail in the following section.

\section{Variance in Fractal Dimension}
\label{sec:variance}

In this section we use the relation between the two point correlation
function, number density of points and the fractal dimension to estimate the
statistical error in determination of the fractal dimension.  
We have shown in \citet{2008MNRAS.390..829B} that for most tracers of the
large scale density field in the universe, the contribution of the finite
number density of points is much smaller than the contribution of clustering
in terms of the deviation of fractal dimension from the ambient dimension.
In the following discussion we assume that the contribution of a finite number
density of points can be ignored.
With this assumption, we elevate the relation between the fractal dimension
and the two point correlation function to the dispersion of the two
quantities. 
The statistical error in the Minkowski-Bouligand Dimension can then be
estimated from the statistical error in the correlation function. 
\begin{equation}
Var\{\Delta D_q\} \simeq Var\{{(\Delta D_q)}_{clus}\}
\end{equation}
This implies that 
\begin{eqnarray}
Var\{\Delta D_q\}&\simeq& Var\{\frac {Dq}{2}\left(\bar\xi(r) -\xi(r) \right)\}
\nonumber \\ 
&=& \left(\frac {Dq}{2}\right)^2 \left( Var\{\bar\xi(r)\} +Var\{\xi(r)\}
  \nonumber \right.\\ 
& &\hspace{1cm}\left.+2Cov\{\bar\xi(r)\xi(r)\}\right) 
\end{eqnarray}
We can make use of the fact that $\left|Cov(x,y)\right| \leq
\sigma_x \sigma_y$ where $x$ and $y$ are random variables.  
We get:
\begin{equation}
\left|{\sigma_{\bar\xi}(r)-\sigma_{\xi}(r)}\right| \leq
\frac{2{\sigma}_{\Delta D_q}}{D q}  \leq
\left|{\sigma_{\bar\xi}(r)+\sigma_{\xi}(r)}\right|     
\end{equation}
If we find that one of the $\sigma_{\bar\xi}(r)$ or $\sigma_{\xi}(r)$ is much
larger than the other then we get:
\begin{equation}
{\sigma}_{\Delta D_q} \simeq \frac{D q}{2} ~
Max\left(\sigma_{\bar\xi}(r),\sigma_{\xi}(r)\right) 
\label{sigdq}
\end{equation}
The problem is then reduced to the evaluation of statistical error in the
correlation function.

Starting point in the analytical estimate of the statistical error in 
$\xi(r)$ is the assumption that the variance in the power spectrum is 
that expected for Gaussian fluctuations with a shot-noise component 
arising from the finite number of objects used to trace the density  field
\citep{1994ApJ...426...23F, 2009MNRAS.400..851S}:   
\begin{equation}
\sigma_{P}(k)=\sqrt{\frac{2}{V}}\left(P(k)+\frac{1}{{\bar n}} \right),
\label{eq:sigmap}
\end{equation}
where $V$ is the simulation volume, and ${\bar n}$ is the mean density of the
objects considered (dark matter particles or
halos). 
\citet{2008MNRAS.383..755A} found good agreement between this
expression and the variance in $P(k)$ measured from numerical simulations.  
In order to develop a consistent approach, we use $1/\bar{n} = 0$ in the
following discussion.  

The covariance of the two-point correlation function is defined by
\citep{1994ApJ...424..569B,2006NewA...11..226C,2008PhRvD..77d3525S}: 
\begin{eqnarray}
 C_{\xi}(r,r')&\equiv&\left\langle (\xi(r)-{\bar \xi}(r))(\xi(r')-{\bar
     \xi}(r'))\right\rangle  \nonumber \\ 
&=&\int \frac{{\rm d}k\,k^2}{2\pi^2}j_0(kr)j_0(kr')\sigma^2_P(k),
\label{eq:cov_xi}
\end{eqnarray}
where the last term can be replaced by Eq.~(\ref{eq:sigmap}).
The variance in the correlation function is simply $\sigma^2_{\xi}(r) =
C_{\xi}(r,r)$. 
\citet{2009arXiv0908.2598K} have tested this formula against the variance
derived from the mock catalogs for both dim and bright galaxy samples of SDSS,
and found that at $50 < r <100 \,h^{-1}$Mpc the variance is consistent with
equation \ref{eq:cov_xi}.   
The direct application of Eq.~(\ref{eq:cov_xi}) leads to a 
substantial over prediction of the variance, since it ignores the effect of
binning in pair separation which reduces the covariance in the measurement
\citep{2006NewA...11..226C, 2008PhRvD..77d3525S}. 

An estimate of the correlation function in the $i^{\rm th}$ pair separation
bin ${\hat \xi}_i$ corresponds to the shell averaged correlation function
\begin{equation}
 {\hat \xi}_i=\frac{1}{V_i}\int_{V_i} \xi(r)\,{\rm d}^3r,
\end{equation}
where $V_i$ is the volume of the shell. 
The covariance of this estimate is given by the following expression, e.g.,
see \citep{2008MNRAS.390.1470S}  
\begin{eqnarray}
C_{{\hat \xi}}(i,j)&=&\frac{1}{V_i\,V_j}\int {\rm d}^3r\int {\rm
  d}^3r'C_{\xi}(r,r') \nonumber \\ 
&=&\int \frac{{\rm d}k\,k^2}{2\pi^2}{\bar j}_0(k,i){\bar j}_0(k,j)\sigma^2_P(k),
\label{eq:cov_xi2}
\end{eqnarray}
where
\begin{equation}
{\bar j}_0(k,i)=\frac{1}{V_i}\int_{V_i} j_0(kr)\,{\rm d}^3r.
\end{equation}
is the volume averaged Bessel function. 
So the variance in $\bar\xi$ is $\sigma^2_{\bar\xi}(r)=C_{\hat \xi}(i,i)$.
We find that at scales of interest $\sigma_{\bar\xi}(r) \ll \sigma_\xi(r)$. 
From here, it is straightforward to compute the standard deviation in ${\Delta
  D_q}$ using equation \ref{sigdq}. 

\begin{figure}
\begin{center}
  \includegraphics[width=2.5in]{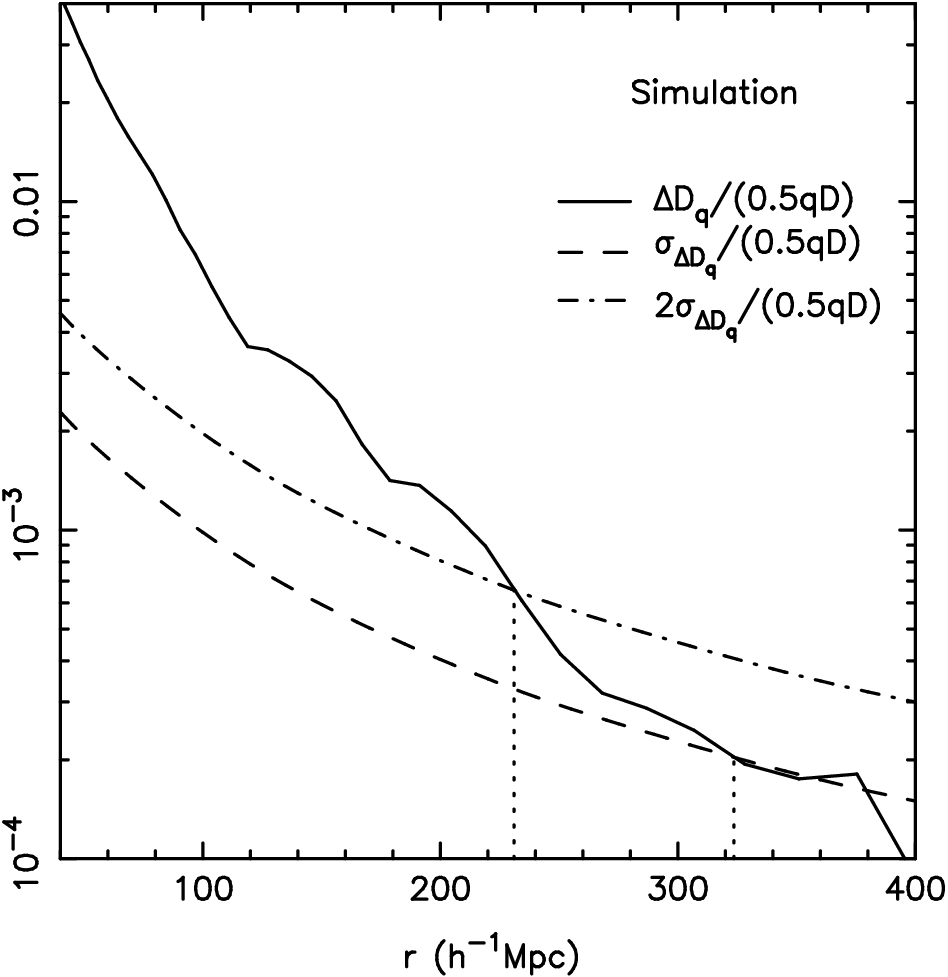}
  \includegraphics[width=2.5in]{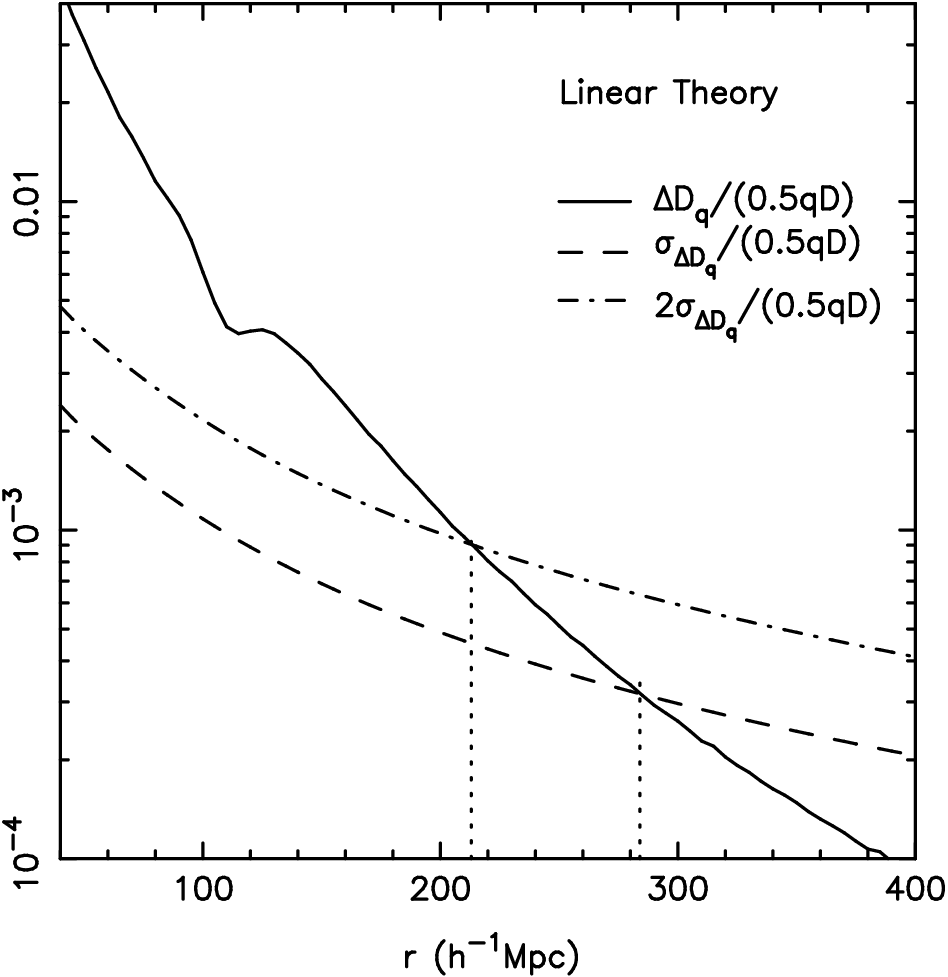}
  \includegraphics[width=2.5in]{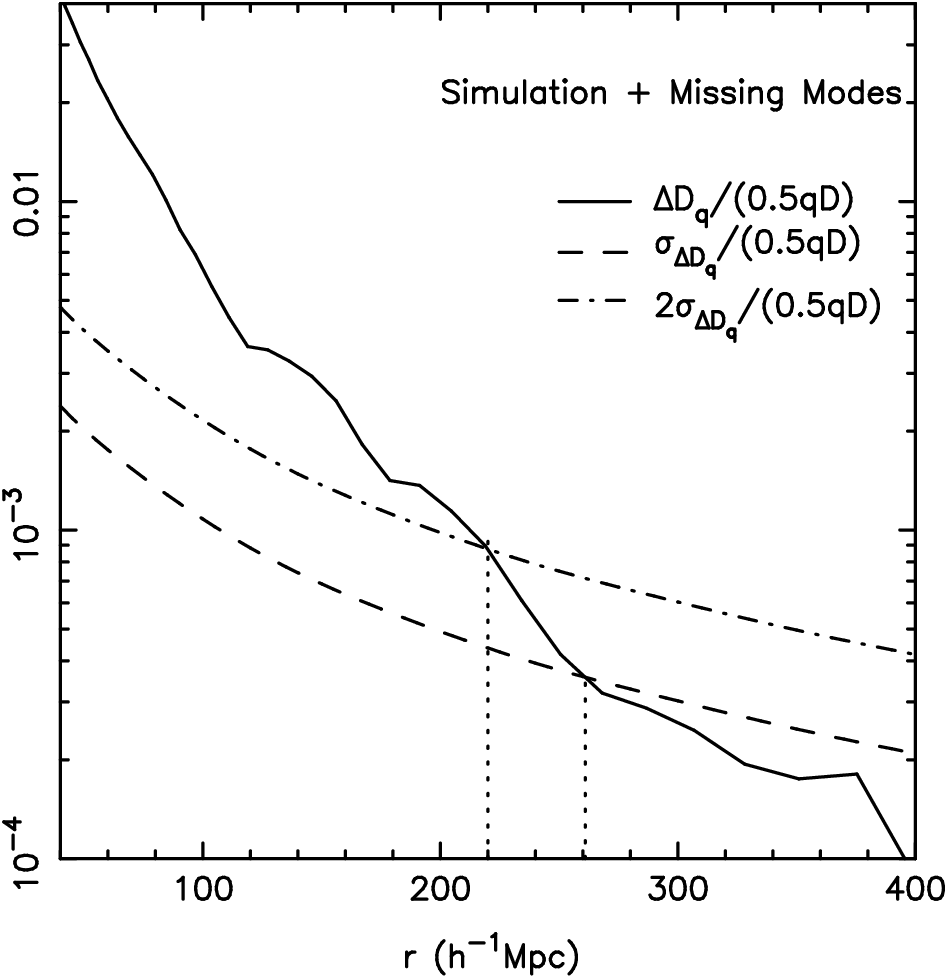}
  \caption{Variation of $\Delta D_q$ and its predicted standard deviation
    with scale is shown in these plots.  The top panel shows these for the 
    $\Lambda$CDM simulation described in the text, the middle panel shows 
    the same using the linearly evolved power spectrum and the lower panel 
    again shows the same quantity with data from simulations patched with 
    the linearly evolved power spectrum at large scales.  $\Delta D_q/(0.5 q
    D)$ is  
    shown using a solid curve in each panel, the dashed line shows the 
    dispersion $\sigma_{\Delta D_q}/(0.5qD)$ and the dot-dashed line shows 
    $2\sigma_{\Delta D_q}/(0.5qD)$.  The intersection of $\Delta D_q/(0.5qD)$
    and $\sigma_{\Delta D_q}/(0.5qD)$ is the scale beyond which we cannot
    distinguish between the $\Lambda$CDM model and a homogeneous distribution.} 
  \label{fig:simdqvsr}
\end{center}
\end{figure}

\section{Scale of Homogeneity}

We can describe a distribution of points to be homogeneous if the standard
deviation of ${\Delta D_q}$ is greater than ${\Delta D_q}$. 
Note that this formulation gives us a unique scale: above this scale it is not
possible to distinguish between the given point distribution and a homogeneous
distribution. 
In the cosmological context, we have bypassed the details of contribution to
the variance arising from the survey geometry, survey size, etc. 
Clearly, if we were to take these contributions into account, the error in
determination of ${\Delta D_q}$ will be larger and hence we will recover a
smaller scale of homogeneity \citep{2009MNRAS.399L.128S}. 
This change results not from any change in the real scale of homogeneity but
because of the limitations of observations.  
In our view, the {\sl real} scale of homogeneity is one where these
limitations do not matter. 
In the limit where we ignore these additional sources of errors, we have the
following general conclusions: 
\begin{itemize}
\item
As long as non-linear correction are not important, the scale of homogeneity
does not change with epoch.
\item
In real space, the scale of homogeneity is independent of the tracer used,
because the deviation ${\Delta D_q}$ as well as the dispersion in this
quantity scale in the same manner with bias. 
This follows from the fact that bias in correlation function can be
approximated by a constant number at a given epoch at sufficiently large
scales\citep{1998MNRAS.299..417B, 2000MNRAS.318..203S}. 
\item
Redshift space distortions introduce some bias dependence in the scale of
homogeneity. 
\item
As long as our assumption of $q\bar\xi \ll 1$ is valid, the scale of
homogeneity is the same for all $q$. 
\end{itemize}
These points highlight the robust and unique nature of the scale of
homogeneity defined in the manner proposed in this work. 

We now turn to more specific conclusions in the cosmological setting below.
We have calculated $\sigma_{\Delta D_q}$ and ${\Delta D_q}$ using the power
spectrum of a large N-body simulation as well as using the linear power
spectrum obtained from WMAP5 parameters. 
We have summarised these findings in Figure~\ref{fig:simdqvsr}.  
The top panel in the figure shows the deviation and the dispersion as
estimated from an N-Body simulation.
In this case the estimated scale of homogeneity is just above
$320$~h$^{-1}$Mpc. 
The corresponding calculation with the linear theory gives a scale of just
over $280$~h$^{-1}$Mpc. 
Part of the reason for this difference is that the simulation does not contain
perturbations at very large scales.  
If we patch the power spectrum derived from simulations with the linearly
extrapolated power spectrum of fluctuations at these scales then we get
$260$~h$^{-1}$Mpc as the scale of homogeneity, broadly consistent with the
value derived from linear theory.  
Curves for this last estimate are shown in the lowest panel of
Figure~\ref{fig:simdqvsr}.  
If we compare $\Delta D_q$ with $2 \sigma_{\Delta D_q}$ instead, then we get
scales of $230$, $215$ and $220$~h$^{-1}$Mpc, respectively, in the three
cases. 
As expected, increasing the error or dispersion leads to a smaller scale of
homogeneity.  
Once again, the answers derived using different approaches are broadly
consistent with each other and even the small deviations can be understood in
terms of generic aspects of non-linear evolution of perturbations
\citep{1996ApJ...472....1B, 1997MNRAS.286.1023B, 2007ApJ...664..660E}.

\section{Summary}

In this paper we have verified the relation between the two point correlation
function and fractal dimensions for a simulated distribution of matter at very
large scales. 
This relation indicates \citep{2008MNRAS.390..829B} that if we use a strongly
biased tracer, such as Luminous Red Galaxies or clusters of galaxies then the
deviation of the fractal dimension from $3$ is larger. 
This is consistent with observations \citep{1997A&AS..123..119E}.
It is also worthwhile to mention that several observations have reported
detection of excess clustering at a scale of $120$~h$^{-1}$Mpc
\citep{1990Natur.343..726B, 1997Natur.385..139E} and distances between the
largest overdensities have been observed to exceed the scale of homogeneity
derived from most observations \citep{2009arXiv0906.5272E}. 

In this paper, we have also used the relation to estimate statistical
uncertainty in determination of fractal dimensions.  
The scale of homogeneity is then taken to be the scale where this uncertainty
is the same as the offset of fractal dimension from $3$.
We show explicitly that the uncertainty scales with bias in the same manner as
the offset of fractal dimension, thus the true scale of homogeneity is not
sensitive to which objects are used as long as the survey volume is large
enough to contain a sufficient number.
The spirit of this paper is to estimate the fractal dimensions for an ideal
observation, and not to worry about the limitations of current observations. 
The connection between the fractal dimension and correlation function allows
us to make this leap in the limit of weak clustering that is clearly
applicable at large scales.
Applying this to a cosmological situation with the model favoured by WMAP-5
observations, we have estimated the scale of homogeneity to be close to
$260$~h$^{-1}$Mpc. 
As we have ignored several sources of uncertainty that are likely to be
present in most observational data sets, this estimated scale of homogeneity
is in some sense the upper limit of what can be determined as the scale of
homogeneity from observations.
It is comforting to note that the scale of homogeneity is much smaller than
the Hubble scale.

An attractive feature of this way of defining the scale of homogeneity is that
it can be defined self consistently within the setting of the cosmological
model with density perturbations. 
Further, this scale is independent of epoch and largely independent of the
choice of tracer for the large scale density field in the universe.
The scale of homogeneity is the same for the entire spectrum of fractal
dimensions, within the constraints of the underlying assumption that
$q\left|\xi\right| \ll 1$, making this a unique scale in the problem. 

A comparison with recent determination of the scale of homogeneity from
observations is pertinent. 
Observational analyses have shown that the scale of homogeneity may be as
small as $60-70$~h$^{-1}$Mpc \citep{2005MNRAS.364..601Y, 2005ApJ...624...54H,
  2009MNRAS.399L.128S}.  
This is much smaller than the scale of homogeneity we have found using our
method. 
However, we have ignored the effect of survey geometry and size, and hence in
the analysis of any observational dataset there are additional contributions
to $\sigma_{\Delta D_q}$.  
Any increase in the value of $\sigma_{\Delta D_q}$ leads to a smaller scale of
homogeneity. 
In this sense our estimate is the {\it ideal} scale of homogeneity and may be
treated as an upper limit. 
Note that the additional contributions to $\sigma_{\Delta D_q}$ are required
to increase its value by more than an order of magnitude above our
determination for the scale of homogeneity to be as small as $70$~h$^{-1}$Mpc.
It should be possible to demonstrate consistency with our calculation through
an explicit estimate of errors in observational survey. 
In the long term we expect that an increase in survey depth and size will
gradually lead to lowering of errors from other sources and we should obtain
a larger scale of homogeneity.  

\section*{Acknowledgments}

Computational work for this study was carried out at the cluster
computing facility in the Harish-Chandra Research Institute
(http://cluster.hri.res.in/).  
This research has made use of NASA's Astrophysics Data System. 
The authors thank Prof. Jaan Einasto for useful suggestions.

\end{document}